\documentclass[twocolumn,showpacs,preprintnumbers,amsmath,pop]{revtex4}

\usepackage{amsmath}
\usepackage{graphicx}
\usepackage{bm}

\begin{document}

\title{Excitation of a global plasma mode by an
intense electron beam in a dc discharge}

\author{D. Sydorenko}

\affiliation{University of Alberta, Edmonton, Alberta T6G 2E1, Canada}

\author{I. D. Kaganovich}

\affiliation{Princeton Plasma Physics Laboratory, Princeton University,
          Princeton, New Jersey 08543, USA}

\author{P. L. G. Ventzek}

\author{L. Chen}

\affiliation{Tokyo Electron America, Inc., Austin, Texas 78741, USA}

\begin{abstract}

Interaction of an intense electron beam with a finite-length, inhomogeneous
plasma is investigated numerically. The plasma density profile is maximal in
the middle and decays towards the plasma edges. Two regimes of the two-stream
instability are observed. In one regime, the frequency of the instability is
the plasma frequency at the density maximum and plasma waves are excited in
the middle of the plasma. In the other regime, the frequency of the
instability matches the local plasma frequency near the edges of the plasma
and the intense plasma oscillations occur near plasma boundaries. The latter
regime appears sporadically and only for strong electron beam currents. This
instability generates copious amount of suprathermal electrons. The energy
transfer to suprathermal electrons is the saturation mechanism of the
instability.

\end{abstract}




\maketitle

\section{Introduction \label{sec:01}}

Beam-plasma instabilities can produce strong electromagnetic fields which
decelerate and scatter beam electrons within distances much shorter than the
collisional mean free path and transfer the beam energy to plasma electrons.
This process is important for wave-particle
interactions,~\cite{TsytovichBook1970,PorkolabRMP1978,ShapiroBPPbook1984,
RobinsonRMP1997} solar disruptions,~\cite{MelroseBook1986} inertial
fusion,~\cite{MalkinPRL2002,KempPRL2006,KempPP2010} collisionless
shocks,~\cite{KeloggPSS2003,TreumannAAR2009,MoritaJPCS2010,BaloghISSISRS2011}
generation of suprathermal electrons,~\cite{YoonPRL2005} \textit{etc}. In
laboratory plasmas it is often necessary to control where and how the electron
beam delivers its energy to the plasma. This is especially important for
advanced plasma processing applications in order to produce high-aspect ratio
nano-features of electronic devices.~\cite{XuAPL2008,ChenPSST2013}

One way to control the beam energy deposition in plasma is by profiling the
background plasma density.~\cite{MalkinPRL2002} Conditions favorable for
development of the beam-plasma instabilities depend on parameters such as the
beam and plasma densities, plasma temperature, and plasma density gradients. By
appropriately controlling the choice of system parameters, it is possible to
produce instabilities and beam energy deposition in desirable regions, while
suppressing instabilities in undesirable regions.

Furthermore, to suppress the Langmuir waves outside the deposition region,
steep plasma density gradients can be formed there to employ convective
stabilization effects.~\cite{RyutovJETP1970} Several experiments indicate that
electron beams can be efficiently transported through inhomogeneous plasmas for
sufficiently steep plasma density gradients. In this case, the beam deposits
its energy in the region of homogenous plasma density.~\cite{GolovanovSJPP1977}

In bounded plasmas, interaction of the beam with the plasma of uniform density
can be described by the modified Pierce
theory.~\cite{KaganovichPP2016,SydorenkoPP2016} This theory predicts that the
amplitude of oscillations grows exponentially along the beam path. Plasma
nonuniformity can completely change the development of the instability.
Simulation of interaction of an electron beam with an inhomogemeous
finite-length plasma discussed in the present paper reveals a different mode
where intense oscillations appear only near the plasma edges, with practically
no plasma wave excited in the middle of plasma. The oscillations at the
opposite boundaries are synchronized. Below, this mode is referred to as the
“global mode”.

Similar regimes have been considered before. It is known that in a finite-size
plasma an eigenmode is excited due to reflection from
boundaries.~\cite{DavidsonBook2001,DavidsonPRSTAB2004} There are a number of
experiments~\cite{WehnerJAP1951,LooneyPR1954,KochmarevPPR1995,HayashiPP1995}
and simulations~\cite{MoreyPFB1989,MatsumotoPP1996} where standing waves are
excited by an electron beam in such a plasma. A possible mechanism of
excitation of the standing wave is as follows. As beam electrons interact with
the plasma wave, the wave modulates the beam velocity when the beam enters the
plasma. Then bunching of the beam electrons takes place while they travel
through the plasma (similar to a klystron); at the exit from the plasma the
electron beam decelerates and transfers its energy to the oscillations. A wave
reflects from the boundary and propagates backwards to the beam injection
location. This provides the feedback between the deceleration and the
modulation/injection areas.~\cite{WehnerJAP1951,LooneyPR1954}

The present paper describes the global mode and studies the mechanism of its
saturation. It is shown that the saturation occurs because of generation of a
large amount of suprathermal electrons. This finding is different from earlier
assumption of Ref.~\onlinecite{MatsumotoPP1996} that the mode saturates due to
the loss of synchronism between the modulated beam and the plasma oscillations
at the beam exit from the plasma, where the oscillations are supposed to gain
their energy from the beam.

A distinct feature of the global mode is the localization of the intense
electric field in the near-wall areas where the density gradient is the
strongest. Previous studies of interaction of a strong, warm electron beam with
an inhomogeneous, bounded plasma showed that the trapped Langmuir waves can
determine the frequencies and position of the wave field even for moderate
density gradients and field strengths below the threshold for Langmuir
collapse.~\cite{GunellPRL1996,McFarlandPRL1998,WendtPS2001} In those studies
the beam energy was relatively low, about 40 eV, and strong acceleration of
plasma electrons was not observed. The present paper demonstrates that
acceleration of plasma electrons can be the main mechanism of saturation of the
instability.

The paper is organized as follows. Section~\ref{sec:02} describes configuration
and parameters of the simulated system. Section~\ref{sec:03} discusses general
properties of the global mode. In Section~\ref{sec:04}, comparison of multiple
rates of energy input and loss is performed. A test particle study of
synchronism between the beam and the global mode is described in
Section~\ref{sec:05}. The concluding remarks are in Section~\ref{sec:06}.

\section{Simulation setup \label{sec:02}}

The study of the global mode is carried out using an electrostatic
particle-in-cell code.~\cite{SydorenkoPhDTh2006} The simulated plasma is
bounded between two electrodes, an anode and a cathode. A one-dimensional
problem is considered, where only the direction $x$ normal to the electrode
surface is resolved. The anode at $x=0$ has a constant positive potential
$U=800\text{~V}$. The cathode at $x=40\text{~mm}$ has a constant zero
potential. The plasma consists of electrons and single-charged argon ions,
there is also a uniform neutral argon gas. Collisions between electrons and
neutrals are included but for the processes considered below they are not
important. The cathode emits a constant flux of electrons.

\begin{figure}[tbp]
\includegraphics {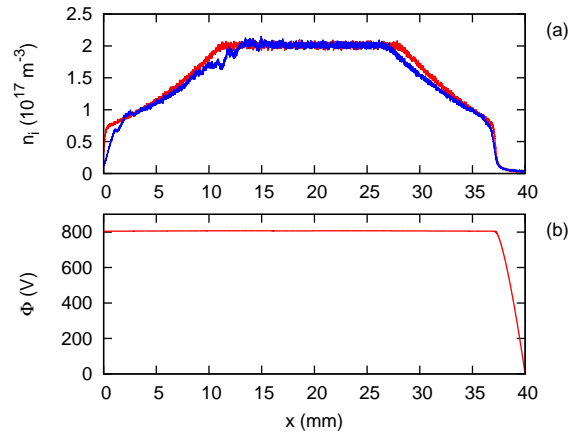}
\caption{\label{fig:01} %
(a) Profiles of ion density at $t=49\text{~ns}$ (red) and $t=499\text{~ns}$
(blue). (b) Profile of the electrostatic potential at $t=49\text{~ns}$. Note
that $t=49\text{~ns}$ is $1\text{~ns}$ before the emission from the cathode
begins. }
\end{figure}

The initial plasma density profile is trapezoidal -- maximal in the center and
decaying towards the plasma edges, which is a frequent situation in a real
discharge plasma. Such a plasma state is obtained as follows. A simulation
starts with a uniform plasma density $n_{e,0}=2\times 10^{17}\text{~m}^{-3}$,
electron temperature $T_{e,0}=2\text{~eV}$, ion temperature
$T_{i,0}=0.03\text{~eV}$, no electron and ion flows, and the electron emission
from the cathode turned off. The plasma evolves for $4000\text{~ns}$, then the
plasma state is saved as the initial state for future simulations. The ion
density and the electrostatic potential profiles in the initial state are shown
by red curves in Fig.~\ref{fig:01}.
%

In the simulation discussed below, the cathode is emitting electrons with the
flux of $\Gamma_{2,e}=5.02\times{10}^{21}\text{m}^{-2}\text{s}^{-1}$, the
corresponding electric current density is $J_e=803.2\text{~A/m}^2$. The ratio
of the beam density and the plasma density in the middle of the plasma is
$n_b/n_{e,0}=0.0015$. This ratio is very small and the effect of the
unmodulated beam charge on plasma oscillations excited in the system is
insignificant. The simulation lasts for $500\text{~ns}$, the emission starts at
$t=50\text{~ns}$.

\section{General properties of the global mode \label{sec:03}}

\begin{figure}[tbp]
\includegraphics {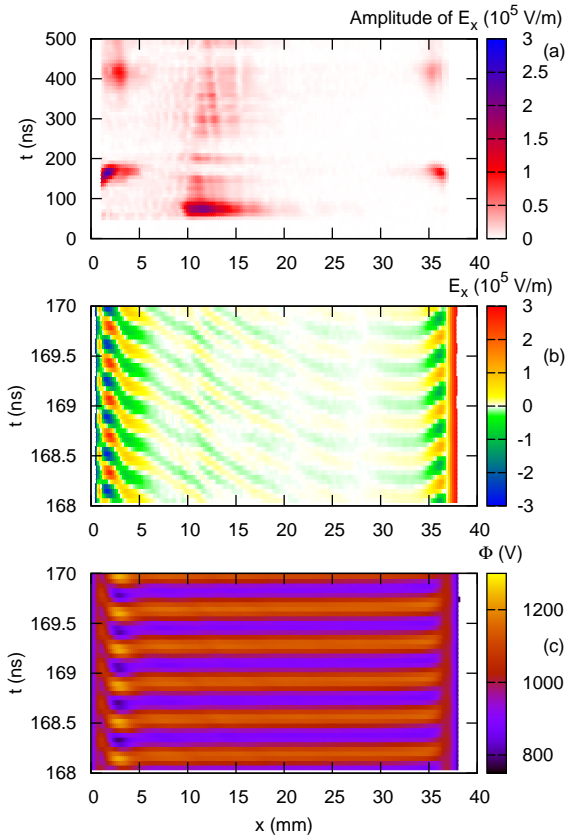}
\caption{\label{fig:02} %
(a) Electric field amplitude of high frequency plasma oscillations vs
coordinate and time.  Electric field (b) and electrostatic potential (c) of the
global mode vs coordinate and time.}
\end{figure}

\begin{figure}[tbp]
\includegraphics {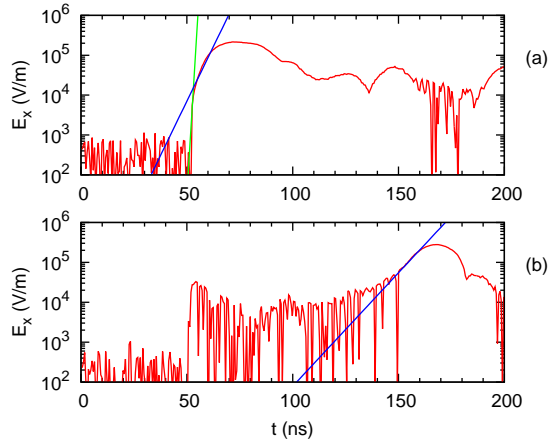}
\caption{\label{fig:03} %
Electric field amplitude of high frequency plasma oscillations vs time at
points with coordinate $x=12\text{~mm}$ (a) and $x=2\text{~mm}$ (b). In (a),
the green curve represents exponential growth of the two-stream instability in
an infinite plasma with the rate of
$0.69(n_b/n_{e,0})^{1/3}\omega_{pe,0}=0.079\omega_{pe,0}$, the blue curve
represents exponential growth with the rate of $0.01\omega_{pe,0}$. In (b), the
blue curve represents exponential growth with the rate of
$0.0052\omega_{pe,0}$. }
\end{figure}

At the very beginning of the simulation ($50\text{~ns}<t<100\text{~ns}$), the
two-stream instability shows an ordinary behavior, see Fig.~\ref{fig:02}(a).
The beam excites oscillations with the frequency equal to the plasma frequency
in the density plateau, the amplitude of the plasma oscillations grows along
the direction of beam propagation, the plasma oscillations quickly decay in the
plasma density gradient areas, and saturation of the instability occurs when
there is strong trapping of both the beam and the bulk electrons. One can
estimate the growth rate of this instability using equation (18) of
Ref.~\onlinecite{KaganovichPP2016}. With $L=17\text{~mm}$ corresponding to the
width of the density plateau, $\alpha\equiv n_b/n_{e,0}=0.0015$, and the beam
velocity corresponding to the 800 V accelerating voltage, one obtains the
growth rate of $0.01\omega_{pe,0}$, where $\omega_{pe,0}=2.52\times
10^{10}\text{~s}^{-1}$ is the plasma frequency in the density plateau region.
Exponential growth with this rate matches well the actual dependence of the
electric field amplitude on time, compare the blue and the red curves in
Fig.~\ref{fig:03}(a). Note that this growth rate is an order of magnitude lower
than the classical growth rate of the two-stream instability in an infinite
plasma, $0.69(n_b/n_{e,0})^{1/3}\omega_{pe,0}=0.079\omega_{pe,0}$, compare the
blue and the green curves in Fig.~\ref{fig:03}(a). The ordinary two-stream
instability is similar to the one considered in
Ref.~\onlinecite{SydorenkoPP2015} and is out of the scope of the present paper.
%

During time interval $130\text{~ns}<t<180\text{~ns}$, intense oscillations
occur near the plasma edges while oscillations in the middle area are much
weaker. The temporal growth rate of these oscillations is about
$0.0052\omega_{pe,0}$ which is lower than that of the ordinary two-stream
instability calculated above. This value is obtained by approximating the
electric field amplitude versus time curve with an exponential law, compare the
red and blue curves in Fig.~\ref{fig:03}(b). Similar regime reappears again at
$390\text{~ns}<t<450\text{~ns}$. This is the so-called global mode. The
intermittent appearance of this mode has been observed by the authors in other
simulations (not shown). An important condition for such a mode is sufficiently
high current of electron emission from the cathode. Below, only the first
occurrence of the global mode is considered since it is characterized by the
highest amplitude.
%

\begin{figure}[tbp]
\includegraphics {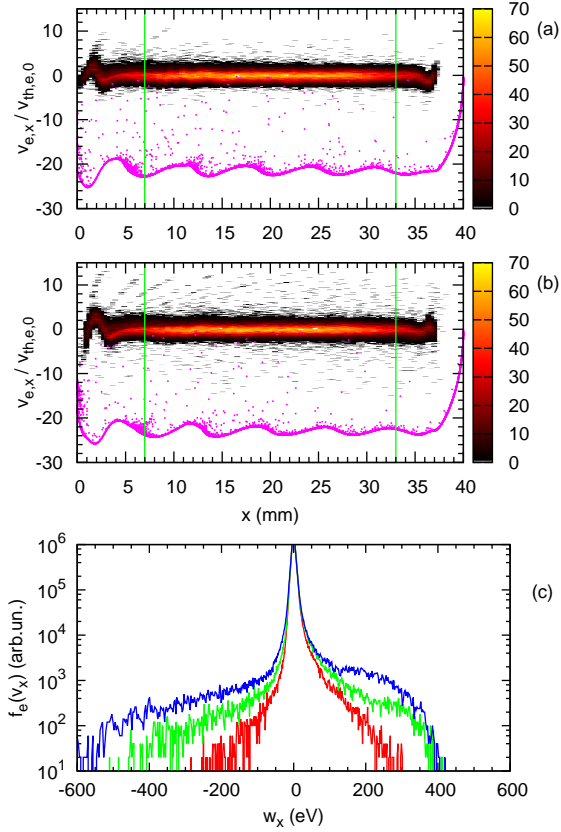}
\caption{\label{fig:04} %
(a,b) Phase planes ``coordinate-velocity'' of the plasma bulk electrons
(colormap) and the beam electrons (magenta) at time $t_A=160\text{~ns}$ (a) and
$t_C=174\text{~ns}$ (b); the scale of the velocity axis is in units of
$v_{th,e,0}=(2T_{e,0}/m_e)^{1/2}$.
(c) The EVDF of bulk electrons at times $t_A$ (red), $t_B=166\text{~ns}$
(green), and $t_C$ (blue); the horizontal coordinate axis is in energy units,
the negative values of the energy correspond to propagation in the negative
$x$-direction.
Times $t_{A,B,C}$ are marked by arrows A, B, and C in
Figure~\ref{fig:06}(a), respectively. }
\end{figure}

This instability is not the ordinary two-stream instability that appears at the
beginning of the simulation. Operation of the plasma beam system in the global
mode resembles operation of a klystron, as suggested in
Ref.~\onlinecite{WehnerJAP1951}. The velocity of the electron beam particles is
modulated by plasma oscillations near the cathode. The amplitude of the
velocity modulation barely grows along the beam up until it approaches the
anode, see Figs.~\ref{fig:04}(a,b). Here the beam transfers its energy to
plasma oscillations. These intense oscillations accelerate bulk electrons to
suprathermal energies, see the enhanced high-energy tails of electron velocity
distribution functions (EVDF) shown in Fig.~\ref{fig:04}(c).
%

The structure of the global mode's electric field is rather complex. In most of
the plasma bulk area there are short waves propagating towards the anode, they
are associated with the density perturbations carried by the modulated beam.
The phase velocity of these waves is equal to the average velocity of the beam.
The short waves are clearly visible in the electric field, see
Fig.~\ref{fig:02}(b). At the same time, Fig.~\ref{fig:02}(c) shows fast
long-wavelength perturbations in the potential. They ensure the feedback
between the oscillations near the anode and the cathode. These probably
correspond to the uniform time dependent component of the electric field that
is part of the Pierce solution~\cite{KaganovichPP2016,SydorenkoPP2016} and
appear to satisfy boundary conditions.

\begin{figure}[tbp]
\centering
\includegraphics {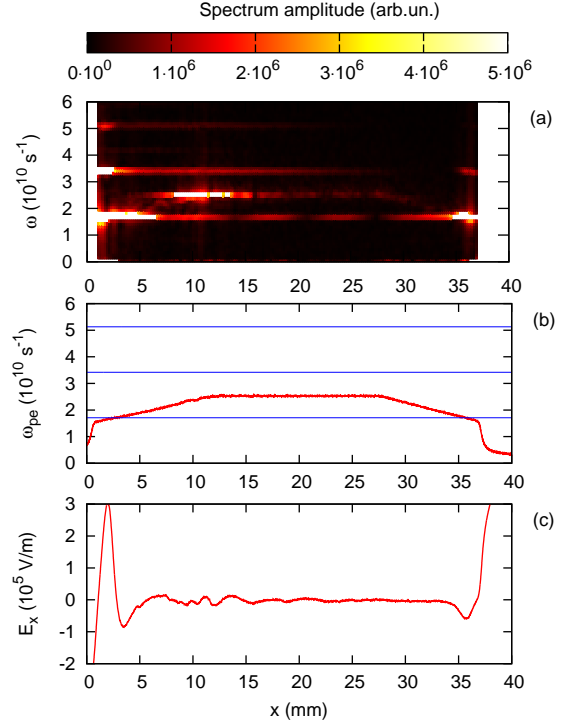}
\caption{\label{fig:05} %
(a) Frequency spectrum of plasma oscillations as a function of coordinate. (b)
Profile of the electron plasma frequency $\omega_{pe}=(n e^2/m_e
\varepsilon_0)^{1/2}$ calculated with the ion density $n=n_i(x,t)$ (red), the
three horizontal blue lines mark harmonics $\omega_0$, $2\omega_0$, and
$3\omega_0$ of the main frequency of the instability $\omega_0=1.71\times
10^{10}\text{~s}^{-1}$. (c) Profile of the electric field. The spectrum in (a)
is obtained for the time interval $154\text{~ns}<t<178\text{~ns}$, the profiles
in (b) and (c) are at time $t=166\text{~ns}$ corresponding to the maximum of
the instability. }
\end{figure}

The frequency with maximal amplitude $\omega_0=1.71\times
10^{10}\text{~s}^{-1}$ is lower than the plasma frequency in the middle of the
plasma $\omega_{pe,0}=2.52\times 10^{10}\text{~s}^{-1}$, see the spectrum in
Fig.~\ref{fig:05}(a). The spectrum also reveals the presence of higher
harmonics of the main frequency $2\omega_0$, $3\omega_0$, etc., and the
frequency of the beam resonance with the plasma in the density plateau
$\omega_{pe,0}$. These frequencies have lower amplitude. The main frequency of
the global mode $\omega_0$ is detected in the spectrum everywhere along the
system. The amplitude of the potential oscillations is maximal where the main
frequency $\omega_0$ is equal to the local electron plasma frequency $\omega_e$
(compare positions of the electric field spikes in Fig.~\ref{fig:05}(c) with
positions of intersections between the plasma frequency profile and the main
instability frequency line in Fig.~\ref{fig:05}(b)).
%

It is necessary to emphasize that such a regime with exactly two resonant
locations $\omega_e(x)=\omega_0$ establishes if (i) there are two areas with
oppositely directed plasma gradients covering same range of plasma frequencies
$\omega_{e,1}<\omega_e<\omega_{e,2}$, (ii) between these areas the plasma
density has a lower limit so that $\omega_e>\omega_{e,2}$, and (iii) outside
these areas the plasma density has an upper limit so that
$\omega_e<\omega_{e,1}$. The electric field of the global mode shown in
Fig.~\ref{fig:02}(b) has the phase of the maximal electric field near the
cathode leading by about $90^\circ$ relative to that near the anode. In this
case, a bunch of beam electrons passes through the strongest decelerating field
near the anode if
%
%
\begin{equation}\label{eq:01}
  n (2\pi v_b/\omega_0)=L_0~,
\end{equation}
where $n$ is a positive integer number, $L_0$ is the distance between points
with $\omega_e=\omega_0$, $v_b$ is the average electron beam velocity, and
$2\pi v_b/\omega_0$ represents the distance between neighbor electron bunches.
Understanding that $L_0$ is a function of the density profile and $\omega_0$,
and that the two points defining the $L_0$ must be taken inside the two density
gradient areas specified above, one may use (\ref{eq:01}) to find possible
values of the global mode frequency $\omega_0$. Note that there may be more
than one solution satisfying $\omega_{e,1}<\omega_0<\omega_{e,2}$ for several
$n$, as well as no solutions at all. For the electron plasma frequency profile
shown in Fig.~\ref{fig:05}(b), one can easily check that equation (\ref{eq:01})
is satisfied for $\omega_0=1.71\times 10^{10}\text{~s}^{-1}$ with $v_b$
corresponding to the bunch energy of 900 eV (plasma potential relative to
cathode at $t=160\text{~ns}$, see Fig.~\ref{fig:06}(b)), $L_b=32.7\text{~mm}$,
and $n=5$.

Finally, it is necessary to mention that the trapezoidal profile shown in
Fig.~\ref{fig:01}(a) is not the only one allowing existence of the global mode.
The authors observed this mode in other simulations (not shown) where the
density profile was non-trapezoidal, with more than one maximum, but still
satisfied the three requirements above.

\section{Energy balance analysis \label{sec:04}}

\begin{figure}[tbp]
\includegraphics {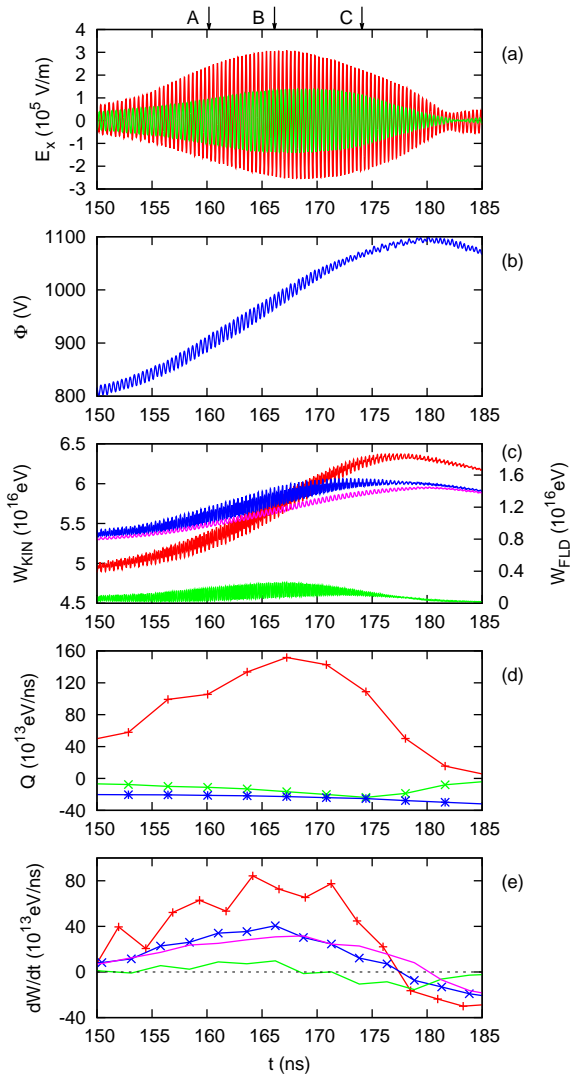}
\caption{\label{fig:06} %
Time dependencies: (a) Electric field at a point with coordinate
$x=2\text{~mm}$ near the anode (red) and at a point with coordinate
$x=36.2\text{~mm}$ near the cathode (green). (b) Electrostatic potential at
$x=36.8\text{~mm}$. (c) Kinetic energy (red, left vertical coordinate axis);
energy of electric field in the whole system (blue), in the sheath areas
(magenta, mostly dc fields), and outside the sheath areas (green, mostly plasma
oscillations), all field-related curves are associated with the right vertical
axis. (d) Rates of beam energy deposition in the system (red), bulk electron
wall losses (green), and ion wall losses (blue); these rates are averaged over
13 oscillation periods. (e) Rates of kinetic energy variation (red), electric
field energy variation in the whole system (blue), sheath electric field energy
(magenta), energy of electric field of plasma oscillations (green)
corresponding to the energy curves in (c) with the same color; these rates are
averaged over 14 oscillation periods. In (a), arrows A, B, and C mark times
when the three EVDFs shown in Fig.~\ref{fig:04}(c) are
obtained. }
\end{figure}

The amplitude of electric field of the global mode both near the cathode and
the anode reaches its maximum at time $167\text{~ns}$, see
Fig.~\ref{fig:06}(a). The growth and subsequent decay of the global mode is
accompanied by the gradual growth of the plasma potential relative to the
cathode, see Fig.~\ref{fig:06}(b). This growth is caused by continuous
acceleration of electrons to suprathermal energies in the near-wall areas where
the electric field of the global mode is maximal, see the electron phase planes
before and after saturation in Figs.~\ref{fig:04}(a) and \ref{fig:04}(b),
respectively.
%

The source of energy for these processes is the anode-cathode voltage $U$. The
rate of energy deposited by the electron beam in the system, $Q_b$, can be
calculated as
\begin{equation}\label{eq:02}
  Q_b=U \Gamma_{2,e} - Q_{b,w}~,
\end{equation}
where $\Gamma_{2,e}$ is the electron flux emitted from the cathode, and
$Q_{b,w}$ is the energy flux carried by the beam electrons across the anode
boundary. The beam energy deposition rate $Q_b$ grows as long as the field
amplitude grows, compare the red curve in Fig.~\ref{fig:06}(d) with
Fig.~\ref{fig:06}(a). This rate stays positive through the instability which
agrees with the finding of Section~\ref{sec:05} that the beam transfers energy
to the global mode all the time. Also, this rate is much higher than rates of
kinetic energy losses due to bulk electrons and ions escaping at the walls, see
the green and the blue curves in Fig.~\ref{fig:06}(d), respectively. Thus, one
can rule out enhanced wall losses as the reason of the saturation.

Both the kinetic and the electric field energies of the system grow through the
instability, see the red and the blue curves in Fig.~\ref{fig:06}(c),
respectively. The rates of growth of these energies are comparable to the rate
of beam energy deposition, compare the red curve in Fig.~\ref{fig:06}(d) with
the red and the blue curves in Fig.~\ref{fig:06}(e). It is reasonable to infer
that the interplay between these three rates has the key to the saturation
mechanism.

Note that after the instability saturates at $t=167\text{~ns}$, the amplitude
of the oscillatory electric field decays while the electric field energy
continues to grow for another 10 ns, compare Fig.~\ref{fig:06}(a) with the blue
curves in Figs.~\ref{fig:06}(c) and~\ref{fig:06}(e). To explain this
difference, it is necessary to consider electric field energy calculated
separately in (i) the near-anode and near-cathode sheath regions
($x<0.6\text{~mm}$ and $x>37\text{~mm}$) and (ii) outside the sheath regions
($0.6\text{~mm}<x<37\text{~mm}$). The former energy associated with the dc
electric field in near-wall regions grows till $t=180\text{~ns}$ corresponding
to the growth of the plasma potential, see the magenta curves in
Figs.~\ref{fig:06}(c) and~\ref{fig:06}(e). The latter energy associated with
the oscillating field of the global mode is much smaller than the former one
and decays after $t=167\text{~ns}$, similar to the instability amplitude, see
the green curves in Figs.~\ref{fig:06}(c) and~\ref{fig:06}(e). Thus, the growth
of the electric field energy in the whole system between 167 ns and 177 ns is
associated with the growth of the sheath dc electric fields.

The kinetic energy increases mostly due to the growth of the number of
suprathermal electrons. The sheath electric field energy also grows because the
most energetic suprathermal electrons escape at the anode. The bulk electrons
are accelerated to suprathermal energies by the intense oscillating fields of
the global mode. Both the kinetic energy and the sheath field energy continue
to grow after the saturation of the instability, when the energy of the global
mode decreases, compare the red (kinetic energy rate) and magenta (sheath field
energy rate) curves with the green curve (global mode field energy rate) in
Fig.~\ref{fig:06}(e). Therefore, it is reasonable to assume that saturation and
decay of the global mode are caused by generation of suprathermal electrons.

It is interesting that while the global mode is the main channel through which
the beam energy is transferred to bulk electrons, the field energy of the
global mode itself is small, see the green curves in Figs.~\ref{fig:06}(c)
and~\ref{fig:06}(e). This means that acceleration of suprathermal electrons is
very efficient in draining the global mode energy. Indeed, the energy of
suprathermal electrons reaches hundreds of eV. Most particles are accelerated
in the strongest field near the anode, see Figs.~\ref{fig:04}(a)
and~\ref{fig:04}(b) for $x<5\text{~mm}$. Acceleration by the intense field near
the cathode becomes noticeable after the saturation of the instability,
probably because it works better for electrons with higher initial energy. Note
that similar effect of enhanced acceleration of particles with higher initial
energy is observed in electron acceleration by plasma waves in nonuniform
plasma density.~\cite{SydorenkoPP2015} One can see ``threads'' of electrons
accelerated near the cathode in a phase plane in Fig.~\ref{fig:04}(b) for
$34\text{~mm}<x<38\text{~mm}$. The result of this additional acceleration is
the asymmetry of the bulk EVDF with significantly higher energy of electrons
flying towards the anode, see Fig.~\ref{fig:04}(c).

\section{Test particle study of beam synchronism \label{sec:05}}

Previously it was pointed out that changes in the regimes of standing waves
excited by electron beams in plasma are associated with breaking of synchronism
between modulation of the beam and the beam energy transfer to the
oscillations.~\cite{MatsumotoPP1996} In order to check whether the saturation
of the global mode occurs because the synchronism between the modulated
electron beam and the plasma oscillations near the anode is disrupted, the
following numerical experiment is carried out. Test electron particles are
emitted from the cathode and travel towards the anode under the influence of
spatially and temporally varying electric fields extracted from the
self-consistent simulation. Modulation of the velocity of test particles occurs
and it results in bunching of the particles.

\begin{figure}[tbp]
\includegraphics {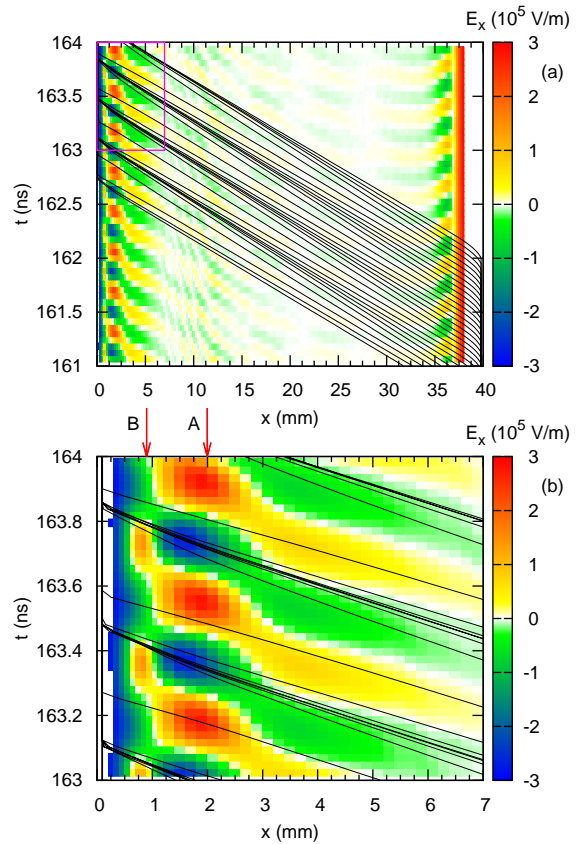}
\caption{\label{fig:07} %
(a) The color map of the electric field versus time and coordinate before the
saturation of the instability. Black curves are trajectories of test particles
injected from the cathode at $x=40\text{~mm}$. Panel (b) is the magnified view
of the area inside the pink box in (a). Arrow A marks location where time
dependencies in Fig.~\ref{fig:09}(b) are obtained. Arrow B marks
location where the beam travels through the accelerating phase of plasma
oscillations. }
\end{figure}

\begin{figure}[tbp]
\includegraphics {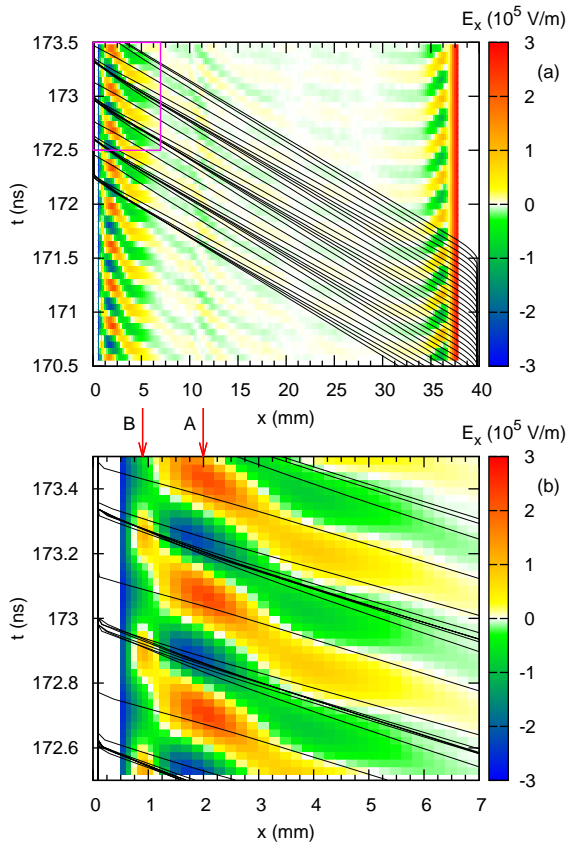}
\caption{\label{fig:08} %
Same as Fig.~\ref{fig:07} for a different time interval, during the decay
stage of the instability. Arrow A marks location where time dependencies in
Fig.~\ref{fig:09}(c) are obtained. }
\end{figure}

It is found that the bunches usually arrive into the decelerating phase of the
electric field oscillations near the anode and therefore amplify the
oscillations. For example, in Fig.~\ref{fig:07} obtained before the global mode
reached saturation, trajectories of most test particles cross the area of
maximal oscillating field ($1\text{~mm}<x<2.4\text{~mm}$) when the electric
field there is negative. A similar picture is observed during the decay of the
global mode, compare Fig.~\ref{fig:08} with Fig.~\ref{fig:07}.
%
%

\begin{figure}[tbp]
\includegraphics {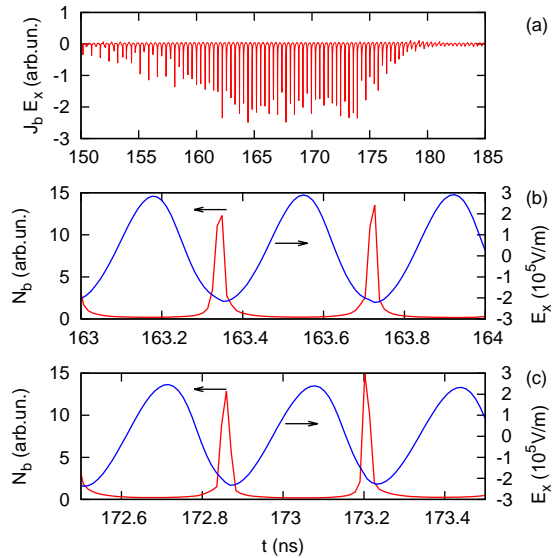}
\caption{\label{fig:09} %
(a) Losses of test particle beam energy at the point with coordinate
$x=2\text{~mm}$ marked by arrow A in Figs.~\ref{fig:07}(b) and \ref{fig:08}(b).
(b) Test particle beam density (red curve, left coordinate axis) and electric
field (blue curve, right coordinate axis) during time interval corresponding to
Fig.~\ref{fig:07}(b). (c) Same as (b) but during time interval corresponding to
Fig.~\ref{fig:08}(b).
}
\end{figure}

For further analysis, location $x=2\text{~mm}$ is selected as a point close to
the center of the near-anode area with maximal oscillating field of the global
mode. In the trajectory plots in Figs.~\ref{fig:07}(b) and \ref{fig:08}(b),
this location is marked with arrow A. During both the growth and the decay
stages, the peak of the test particle beam density occurs very close to the
negative maximum of the electric field (which is the decelerating field for
electrons coming from the cathode), see Figs.~\ref{fig:09}(b) and
\ref{fig:09}(c). In fact, during virtually the whole instability interval, the
product of the test particle beam current and the electric field at this
location is negative, corresponding to energy transfer from the beam to the
wave, see Fig.~\ref{fig:09}(a).
%

It is useful to mention that there is another maximum of the oscillatory field
near the anode, near $x=0.8\text{~mm}$, marked by arrow B in
Figs.~\ref{fig:07}(b) and \ref{fig:08}(b). The electric field there has
opposite direction compared to the main peak marked by arrow A. As a result,
the beam bunches cross this peak when the electric field is accelerating. This
means that the beam takes back some energy from the field. However, the field
amplitude there is lower than in the main peak and the width is smaller as
well, see Figs.~\ref{fig:07}(b) and \ref{fig:08}(b). Therefore, it is safe to
assume that the effect of this small field is minor.

Thus, in contrast to what was suggested in Ref.~\onlinecite{MatsumotoPP1996},
the test particle study shows that in the case considered, breaking of the
synchronism between the beam and the wave is not the reason for the saturation
of the global mode. The breaking of the synchronism does not occur as long as
equation (\ref{eq:01}) has a solution for the given integer number $n$. It is
the gradient of the density which allows the system to find such a solution
when the velocity of electron bunches $v_b$ increases due to the growth of the
plasma potential. In simulation, the mode adjusts its frequency by gradually
shifting the intense field areas closer to each other, see
Fig.~\ref{fig:02}(a), which increases local values of plasma density and
frequency, see Fig.~\ref{fig:05}(b). The frequency adjustment does not have to
be large. For example, the beam phase planes shown in Fig.~\ref{fig:04}(a) and
Fig.~\ref{fig:04}(b) both show five bunches within the gap between the intense
high-frequency field areas, corresponding to $n=5$ in (\ref{eq:01}), while
there is 150 Volts difference in the plasma potential. One can easily estimate
that this requires increase of the mode frequency by less than one percent.

\section{Conclusions \label{sec:06}}

The global mode considered in this paper is an interesting regime when intense
plasma oscillations at the opposite plasma boundaries are synchronized and the
beam-plasma system operates similar to a klystron. This regime is intermittent,
with on and off periods. Sometimes it co-exists with the ordinary two-stream
instability, sometimes it completely replaces it. The global mode requires
relatively high beam current, which is more than $800\text{~A/m}^2$ in the
simulation considered above. Simulations with few times lower current and
otherwise similar initial state do not reveal this mode, see
Ref.~\onlinecite{SydorenkoPP2015}.

A detailed study of the intermittency of the global mode is beyond the scope of
the present paper. One possible reason may be associated with variation of the
density profile which creates or destroys favorable conditions for the
instability defined by equation (\ref{eq:01}). It is necessary to mention that
three major factors affecting the density profile in this study are (a)
acceleration of ions towards the walls due to the ambipolar electric field
which shrinks the plateau and reduces the density elsewhere, (b) escape of ions
at the cathode which reduces the width of the plasma slab, and (c)
ponderomotive effect of intense high-frequency electric fields which creates
localized perturbations with short spatial scale. The density profile in the
end of the simulation formed as a result of these factors is shown by the blue
curve in Fig.~\ref{fig:01}(a).

Although the global mode appears for short periods, it is very efficient in
producing suprathermal electrons, which may be critical for some applications.
This paper finds that generation of the suprathermal electrons is the actual
mechanism of saturation of the global mode, rather than previously suggested
disruption of synchronism between the beam and the wave.

\section*{ACKNOWLEDGMENTS}

D.~Sydorenko and I.~D.~Kaganovich are supported by the U.S. Department of
Energy.

%

%

\end{document}